\def\lb{\hfil\break}
\def\pmb#1{\setbox0=\hbox{#1}%
   \kern-.025em\copy0\kern-\wd0
   \kern.05em\copy0\kern-\wd0
   \kern-0.025em\raise.0433em\box0}
\def\gta{\mathrel{{\lower 3pt\hbox{$\mathchar"218$}}\hskip-8pt
   \raise 2pt\hbox{$\mathchar"13E$}}}
\def\lta{\mathrel{{\lower 3pt\hbox{$\mathchar"218$}}\hskip-8pt
   \raise 2pt\hbox{$\mathchar"13C$}}}
\def\half{{\scriptstyle{1\over2}}}
\def\dagg{\phantom{\dagger}}            
\def\subboldc{\pmb{$\scriptstyle c$}}   
\def\boldd{\pmb{d}}                     
\def\boldell{\pmb{$\ell$}}              
\def\subboldell{\pmb{$\scriptstyle\ell$}}
\def\boldT{\pmb{$T$}}                   
\def\bolddelta{\pmb{$\delta$}}
\def\subbolddelta{\pmb{$\scriptstyle\delta$}}
\def\boldeta{\pmb{$\eta$}}
\def\subboldeta{\pmb{$\scriptstyle\eta$}}
\def\boldxi{\pmb{$\xi$}}
\def\vacuum{|\pmb{\it O}\thinspace\rangle}
\def\bfnabla{\bf \pmb{$\nabla$}\/}
\def\bfcalE{\bf \pmb{${\cal E}$}\/}
\def\up{\uparrow}
\def\dn{\downarrow}
\def\ud{\uparrow\downarrow}
\def\today{\number\day\space\ifcase\month\or
  January\or February\or March\or April\or May\or June\or
  July\or August\or September\or October\or November\or December\fi
 \space\number\year}
\documentstyle[twocolumn,aps]{revtex}
\font\tenrm=cmr8
\begin{document}
\newcommand{\Vv }{{\raisebox{-1.2pt}{\makebox(0,0){$o$}}}}
\newcommand{\Zz }{{\raisebox{-1.2pt}{\makebox(0,0){$\mbox{\tiny o}$}}}}
\newcommand{\Xx }{{\special{em:moveto}}}
\newcommand{\Yy }{{\special{em:lineto}}}
\newcommand{\Ww }{{\usebox{\plotpoint}}}
\title{\begin{minipage}{480pt}
 {\huge \ \\ \ \\ \ \\  \ \\
Stripes, Electron-Like and Polaron-Like Carriers, \\
and High-$T_c$ in the Cuprates \\
\ \\}
{\large \bf J. Ashkenazi$^1$\ \\ }
\end{minipage}
}
\author{\ \hfill
\begin{minipage}{5in}
\small
\baselineskip 11pt
\underline{\it Received 4 November 1998 \ \ \ \ \ \ \ \ \ \ \ \ \ \ \
\ \ \ \ \ \ \ \ \ \ \ \ \ \ \ \ \ \ \ \ \ \ \ \ \ \ \ \ \ \ \ \ \ \ \ \
\ \ \ \ \ \ \ \ \ \ \ \ \ \ \ \ \ \ \ \ \ \ \ \ \ } \\
Both ``large-$U$'' and ``small-$U$'' orbitals are used to study the
electronic structure of the high-$T_c$ cuprates. A striped structure
with three types of carriers are induced, polaron-like ``stripons''
which carry charge, ``quasielectrons'' which carry both charge and spin,
and ``svivons'' which carry spin and lattice distortion. Anomalous 
physical properties of the cuprates are derived, and specifically 
the systematic behavior of the resistivity, Hall constant, and
thermoelectric power. Transitions between
\underline{pair states of quasielectrons and stripons drive
high-temperature superconductivity. \ \ \ \ \ \ } \\ 
{\footnotesize {\bf KEY WORDS:} high-$T_c$ superconductivity, stripes, 
transport properties, mechanism.}
\end{minipage}
}
\maketitle
\setlength{\unitlength}{1in}
\footnotetext[1]{\footnotesize Physics Department, University of Miami,
P.O. Box 248046, Coral Gables, FL 33124, U.S.A.} 
\makeatletter
\global\@specialpagefalse
\def\@oddhead{\footnotesize \it Journal of Superconductivity,
Vol. ??, No. ?, 1999 \hfill \ }
\makeatother
\baselineskip 12pt
\normalsize \rm
\par\noindent{\bf 1. INTRODUCTION}
\medskip

Evidence is growing [1] that the CuO$_2$ planes in the high-$T_c$
cuprates possess a static or a dynamic striped structure. The physical
properties of these materials, and specifically their transport
properties are characterized by intriguing anomalies suggesting the
inadequacy of the Fermi-liquid scenario, and/or the coexistence of
itinerant and almost localized (or polaron-like) carriers. 

First-principles electronic structure studies [2] suggest that realistic
theoretical models of the electrons in the vicinity of the Fermi level
($E_{_{\rm F}}$) should take into account both ``large-$U$'' and
``small-$U$'' orbitals [3].  Let us denote the fermion creation operator
of a small-$U$ electron in band $\nu$, spin $\sigma$, and wave vector
${\bf k}$ by $c_{\nu\sigma}^{\dagger}({\bf k})$.

\bigskip
\par\noindent{\bf 2. AUXILIARY SPACE}
\medskip

Let us treat the large-$U$ orbitals by the ``slave-fermion'' method [4].
A large-$U$ electron in site $i$ and spin $\sigma$ is then created by
$d_{i\sigma}^{\dagger} = e_i^{\dagger} s_{i,-\sigma}^{\dagg}$, if it is
in the ``upper-Hubbard-band'', and by $d_{i\sigma}^{\prime\dagger} =
\sigma s_{i\sigma}^{\dagger} h_i^{\dagg}$, if it is in a Zhang-Rice-type
``lower-Hubbard-band''. Here $e_i^{\dagg}$ and $h_i^{\dagg}$ are
(``excession'' and ``holon'') fermion operators, and
$s_{i\sigma}^{\dagg}$ are (``spinon'') boson operators. The constraint: 
\begin{equation}
e_i^{\dagger} e_i^{\dagg} + h_i^{\dagger} h_i^{\dagg} + \sum_{\sigma}
s_{i\sigma}^{\dagger} s_{i\sigma}^{\dagg} = 1 
\end{equation}
should be satisfied in every site. 

Within an auxiliary Hilbert space a chemical-potential-like Lagrange
multiplier is introduced to impose the constraint on the average.
Physical observables are then projected into the physical space by
taking appropriate combinations of Green's functions of the auxiliary
space. Since the time evolution of Green's functions is determined by
the Hamiltonian which obeys the constraint rigorously, effects of
constraint violation may result only from the approximations applied
treating these Green's functions. {\it E.g.}, within the ``spin-charge
separation'' approximation two-particle spinon-holon Green's functions
are decoupled into products of one-(auxiliary)-particle Green's
functions. 

The spinon states are diagonalized by applying the Bogoliubov
transformation: 
\begin{equation}
s_{\sigma}^{\dagg}({\bf k}) = \cosh{(\xi_{\sigma{\bf k}})}
\zeta_{\sigma}^{\dagg}({\bf k}) + \sinh{(\xi_{\sigma{\bf k}})}
\zeta_{-\sigma}^{\dagger}(-{\bf k}). 
\end{equation} 
The Bose operators $\zeta_{\sigma}^{\dagger}({\bf k})$ create spinon
states with ``bare'' energies $\epsilon^{\zeta} ({\bf k})$ which have a
V-shape zero minimum at ${\bf k}={\bf k}_0$, whose value is either $(
{\pi \over 2{\rm a}} , {\pi \over 2{\rm a}} )$ or $( {\pi \over 2{\rm
a}} , -{\pi \over 2{\rm a}} )$. Bose condensation results in
antiferromagnetism (AF), and the spinon reciprocal lattice is extended
by adding the wave vector ${\bf Q}=2{\bf k}_0$. 

The slave-fermion method is known to describe well an AF state. Since
within this method AF order is obtained by the Bose condensation of
spinons, the decoupling of two-particle spinon-spinon Green's functions,
relevant for physical spin processes, does not harm the treatment of
spin-spin correlations. 

\bigskip
\par\noindent{\bf 3. STRIPES AND QUASIPARTICLES}
\medskip

Theoretically, a lightly doped AF plane tends to separate into a
``charged'' phase and an AF phase. A preferred structure under long-range
Coulomb repulsion is [5] of frustrated stripes of these phases.
Experiment [6]  confirms such a scenario, and neutron-scattering
measurements [7] indicate at least in certain cases a structure where
narrow charged stripes form antiphase domain walls separating wider AF
stripes. Growing evidence [1] supports the assumption that such a
structure exists, at least dynamically, in all the superconducting
cuprates. 

Since the spin-charge separation approximation is valid in
one-dimension, it should apply for holons (excessions) within the
charged stripes, and they are referred to as ``stripons''. They carry
charge, but not spin. We denote their fermion creation operators by
$p^{\dagger}_{\mu}({\bf k})$, and their bare energies by
$\epsilon^p_{\mu}({\bf k})$.

It is evident [1] that the stripes in the cuprates are far from being
perfect. Even when they are not dynamic, one expects them to be defected
and ``frustrated'', and to consist of disconnected segments. Such a
structure is fatal for itinerancy in one-dimension, and it is reasonable
to choose localized states as the starting point for the stripon states.

Other carriers (of both charge and spin) result from the hybridization
(in the auxiliary space) of small-$U$ electrons and coupled
holon-spinons (excession-spinons) within the AF stripes. We refer to 
these carriers as ``Quasi-electrons'' (QE's), and denote their fermion 
creation operators by $q_{\iota\sigma}^{\dagger}({\bf k})$. Their bare
energies $\epsilon^q_{\iota} ({\bf k})$ form quasi-continuous ranges of
bands crossing $E_{_{\rm F}}$ over ranges of the Brillouin zone (BZ). 

\makeatletter
\global\@specialpagefalse
\def\@oddhead{\ \\ \ }
\makeatother

\bigskip
\par\noindent{\bf 4. SPECTRAL FUNCTIONS}
\medskip

The physical observables are evaluated using electrons Green's
functions. Expressions are derived where the observables are expressed
in terms of the auxiliary space spectral functions $A^q_{\iota}({\bf k},
\omega)$, $A^{\zeta}_{\lambda}({\bf k}, \omega)$, and $A^p_{\mu}({\bf
k}, \omega)$, of the QE's, spinons, and stripons, respectively. 

The quasiparticle fields are strongly coupled to each other due to
hopping and hybridization terms of the Hamiltonian. This coupling can be
expressed through an effective Hamiltonian term whose parameters can be
in principle derived self-consistently from the original Hamiltonian. It
has the form:
\begin{eqnarray}
{\cal H}^{\prime} &=& {1 \over \sqrt{N}} \sum_{\iota\mu\lambda\sigma}
\sum_{{\bf k}, {\bf k}^{\prime}} \Big\{\sigma
\epsilon^{qp}_{\iota\mu\lambda\sigma}({\bf k}^{\prime}, {\bf k})
q_{\iota\sigma}^{\dagger}({\bf k}) p_{\mu}^{\dagg}({\bf k}^{\prime})
\nonumber \\ &\ &\times\big[ \cosh{(\xi_{\lambda\sigma,({\bf k} - {\bf
k}^{\prime})})} \zeta_{\lambda\sigma}^{\dagg}({\bf k} - {\bf
k}^{\prime}) \nonumber \\ &\ &+ \sinh{(\xi_{\lambda\sigma,({\bf k} -
{\bf k}^{\prime})})} \zeta_{\lambda,-\sigma}^{\dagger}({\bf k}^{\prime}
- {\bf k}) \big] + h.c. \Big\}, 
\end{eqnarray} 

The auxiliary space spectral functions are calculated through the
standard diagrammatic technique where ${\cal H}^{\prime}$ introduces a
vertex connecting QE, stripon and spinon propagators. It turns out that
the stripon bandwidth is at least an order of magnitude smaller than the
QE and spinon bandwidths. Thus, by a generalized Migdal theorem one gets
that ``vertex corrections'' are negligible, and a second-order
perturbation expansion in ${\cal H}^{\prime}$ is applicable. 

Applying the diagrammatic technique, self-consistent expressions are
derived for the scattering rates $\Gamma^q_{\iota}({\bf k}, \omega)$,
$\Gamma^{\zeta}_{\lambda}({\bf k}, \omega)$, and $\Gamma^p_{\mu}({\bf
k}, \omega)$, of the QE's, spinons, and stripons, respectively. For
sufficiently doped cuprates the self-consistent solution has the
following features: 

\medskip
\par\noindent\underbar{Spinons}

The spinon spectral functions behave as: $A^{\zeta}({\bf k},
\omega)\propto\omega$ for small $\omega$. Thus $A^{\zeta}({\bf k},
\omega) b_{_T}(\omega)\propto T$ for $\omega\ll T$, where
$b_{_T}(\omega)$ is the Bose distribution function at temperature $T$.
Namely there is no long-range AF order (associated with the divergence
in the number of spinons at ${\bf k}={\bf k}_0$). 

\medskip
\par\noindent\underbar{Stripons}

The coupling between the stripon field and the other fields results in 
the renormalization of the localized stripon energies into a very narrow
range around $E_{_{\rm F}}$ (thus getting polaron-like states). Some
hopping via QE-spinon states results is the onset of itineracy at low
temperatures, with a bandwidth of $\sim$$0.02\;$eV. The stripon
reciprocal lattice is extended by adding wave vectors corresponding to
the approximate periodicity of the striped structure. 

The stripon scattering rates can be expressed as; 
\begin{equation}
\Gamma^p({\bf k}, \omega) \propto A \omega^2 + B \omega T + CT^2. 
\end{equation}

\par\noindent\underbar{Quasi-electrons}

The QE scattering rates, resulting from their coupling to the other
fields, can be approximately expressed as: 
\begin{equation}
\Gamma^q({\bf k}, \omega) \propto \omega[b_{_T}(\omega) + \half]. 
\end{equation}
This  becomes $\Gamma^q({\bf k}, \omega)\propto T$ in the limit $T\gg
|\omega|$, and $\Gamma^q({\bf k}, \omega)\propto\half |\omega|$ in the
limit $T\ll |\omega|$, in agreement with ``marginal Fermi liquid''
phenomenology [8]. 

\bigskip
\par\noindent{\bf 5. SOME PHYSICAL ANOMALIES}
\bigskip

\par\noindent{\bf 5.1 Lattice effects (``svivons'')}
\medskip

As was found by Bianconi {\it et al.} [6] the charged stripes are
characterized by LTT structure, while the AF stripes are characterized
by LTO structure. Thus, in any physical process induced by ${\cal
H}^{\prime}$ (3), where a stripon transforms into a QE, or vice versa,
followed by the emission/absorption of a spinon, phonons are also
emitted/absorbed, and the stripons have lattice features of polarons. 

The result is that a spinon propagator linked to the ${\cal H}^{\prime}$
vertex is ``dressed'' by phonon propagators. We refer to such a
phonon-dressed spinon as a svivon. Its propagator can be expressed as a
spinon propagator multiplied by a power series of phonon propagators. As
dressed spinons the svivons carry spin, but not charge; however they
also ``carry'' lattice distortion. 

\bigskip
\par\noindent{\bf 5.2 Optical conductivity}
\medskip

The optical conductivity of the doped cuprates is characterized [9] by a
Drude term and mid-IR peaks. Transitions between low energy QE states
result in the Drude term. Excitations of the very low energy stripon
states result in the mid-IR peaks. Such excitation can either leave a
stripon in the same stripe segment, exciting spinon and phonon states,
or transform a stripon into a QE-svivon state. 

\bigskip
\par\noindent{\bf 5.3 Spectroscopic anomalies}
\medskip

The electronic spectral function, measured, {\it e.g.}, in photoemission
experiments, includes ``coherent'' bands, and an ``incoherent''
background of a comparable integrated weight. The coherent part is due 
to few QE bands, and the frequently observed $\sim$$|E-E_{_{\rm F}}|$ 
bandwidth is consistent with Eq. (5). The incoherent part is due to the
quasi-continuum of other QE bands, and to the stripon states (note that
the spectroscopic signature of stripons is smeared over few tenths of an
eV around $E_{_{\rm F}}$ due to the accompanying svivon excitations).

The observed ``Shadow bands'', ``extended'' van Hove singularities
(vHs), and normal-state pseudogaps result from the effect of the striped
structure on the QE bands [10], due to the extension of the reciprocal
lattice (discussed above). The vHs are extended to supply the spectral
weight for the stripon states, and when the vHs are missing this
transferred spectral weight causes a pseudogap in the same place in the
BZ [11]. 

\bigskip
\par\noindent{\bf 6. TRANSPORT PROPERTIES}
\medskip

\par\noindent{\bf 6.1 Electric current (dc)}
\medskip

The electric current ${\bf j}$ is expressed as a sum of QE and stripon
contributions ${\bf j}^q$ and ${\bf j}^p$, respectively: 
\begin{equation}
{\bf j} = {\bf j}^q + {\bf j}^p.
\end{equation}
As was discussed above, stripons transport occurs through transitions to
intermediate QE-spinon states. Consequently, the expressions for the
currents yield: 
\begin{equation}
{\bf j}^p \cong \alpha {\bf j}^q, 
\end{equation}
where $\alpha$ is approximately $T$-independent. This condition is
satisfied by the formation of gradients $\bfnabla\mu^q$ and
$\bfnabla\mu^p$ of the QE and stripon chemical potentials (respectively)
in the presence of an electric field. The constraint on the number of
electrons imposes: 
\begin{equation}
N^q\bfnabla\mu^q + N^p\bfnabla\mu^p = 0,
\end{equation}
where $N^q$ and $N^p$ are, respectively, the contributions of QE's and
stripons to the electrons density of states at $E_{_{\rm F}}$. 

\bigskip
\par\noindent{\bf 6.2 Electrical conductivity and Hall constant}
\medskip

The Kubo formalism is applied to derive expressions for the dc
conductivity and Hall constant in terms of Green's functions. Such 
expressions are based on diagonal and non-diagonal conductivity QE terms
$\sigma_{xx}^{qq}$ and $\sigma_{xy}^{qqq}$, stripon terms
$\sigma_{xx}^{pp}$ and $\sigma_{xy}^{ppp}$, and mixed terms
$\sigma_{xy}^{qqpp}$. 

The currents in an electric field ${\bf E}$ are expressed as: 
\begin{equation}
j_x^q=\sigma_{xx}^{qq} {\cal E}_x^q, \ \ \ \ \ j_x^p=\sigma_{xx}^{pp}
{\cal E}_x^p, 
\end{equation}
where
\begin{equation}
\bfcalE^q={\bf E} + {\bfnabla\mu^q \over {\rm e}}, \ \ \ \ \
\bfcalE^p={\bf E} + {\bfnabla\mu^p \over {\rm e}}. 
\end{equation}

Using Eqs. (8), (10), one can express:
\begin{equation}
{\bf E}={N^q \bfcalE^q + N^p\bfcalE^p \over N^q+N^p}, 
\end{equation}
and since 
\begin{equation}
j^q_x + j^p_x = j_x = {E_x \over \rho_x}, 
\end{equation}
the resistivity in the $x$- direction $\rho_x$ can be expressed as: 
\begin{equation}
\rho_x = {1 \over (N^q+N^p) (1+\alpha)} \Big( {N^q \over
\sigma_{xx}^{qq}} + {\alpha N^p \over \sigma_{xx}^{pp}}\Big).  
\end{equation}

Similarly, the Hall constant $R_{_{\rm H}} = E_y/j_xH$ can be 
expressed as
\begin{equation}
R_{_{\rm H}} = {\rho_x \over \cot{\theta_{_{\rm H}}}}, 
\end{equation}
where
\begin{equation}
{1+\alpha \over \cot{\theta_{_{\rm H}}}} = {\sigma_{xy}^{qqq} +
\sigma_{xy}^{qqpp} \over \sigma_{xx}^{qq}} + {\alpha(\sigma_{xy}^{ppp} +
\sigma_{xy}^{qqpp}) \over \sigma_{xx}^{pp}}. 
\end{equation}

\vskip -1.1truecm
\begin{figure}[t]
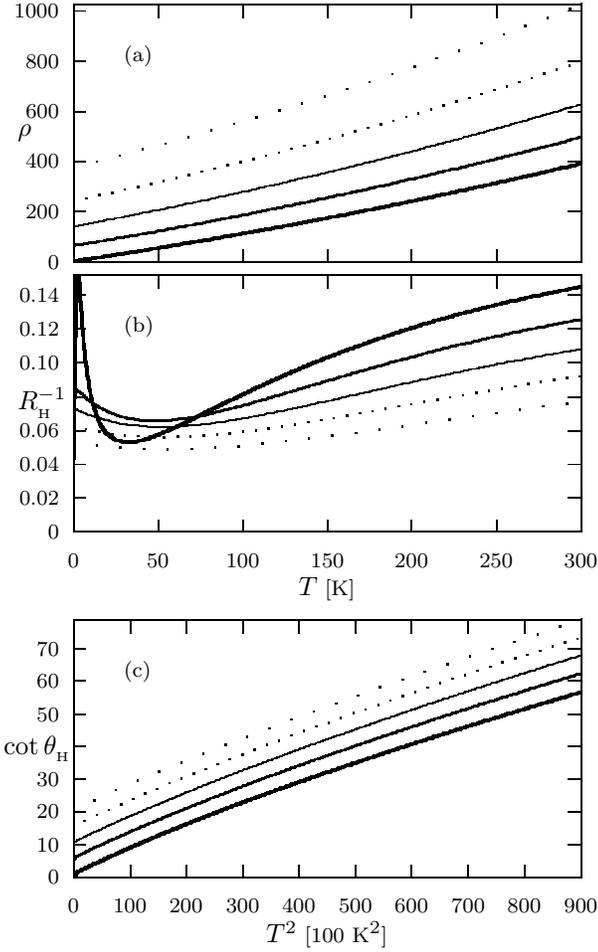


\setlength{\unitlength}{0.240900pt}
\ifx\plotpoint\undefined\newsavebox{\plotpoint}\fi
\sbox{\plotpoint}{\rule[-0.175pt]{0.350pt}{0.350pt}}%


\caption{The resistivity (a), inverse Hall constant (b), and
$\cot{\theta_{_{\rm H}}}$ (c), in arbitrary unit, for parameter values:
A=1,7,13,19,25; B=.001; C=1; D=0,50,100,150,200; N=1,.9,.8,.7,.6; Z=2.
The first value corresponds to the thickest lines.} 
\label{F1}
\end{figure}

In order to find the temperature dependencies of the transport
quantities we apply our results for the scattering rates $\Gamma^q$ and
$\Gamma^p$, given in Eqs. (4), (5), to which we add
temperature-independent impurity scattering terms. Consequently one can
express the temperature dependencies of the conductivity terms using
parameters $A$, $B$, $C$, $D$, $N$, and $Z$, as follows: 
\begin{eqnarray}
\sigma_{xx}^{qq} &\propto& {1 \over D+CT}, \\ \sigma_{xx}^{pp} &\propto&
{1 \over A+BT^2}, \\  \sigma_{xy}^{qqq} &\propto& {1 \over (D+CT)^2}, \\ 
\sigma_{xy}^{ppp} &\propto& {1 \over (A+BT^2)^2}, \\ \sigma_{xy}^{qqpp} 
&\propto& {1 \over (D+CT)(A+BT^2)}. 
\end{eqnarray} 

\vskip -1.1truecm
\begin{figure}[t]
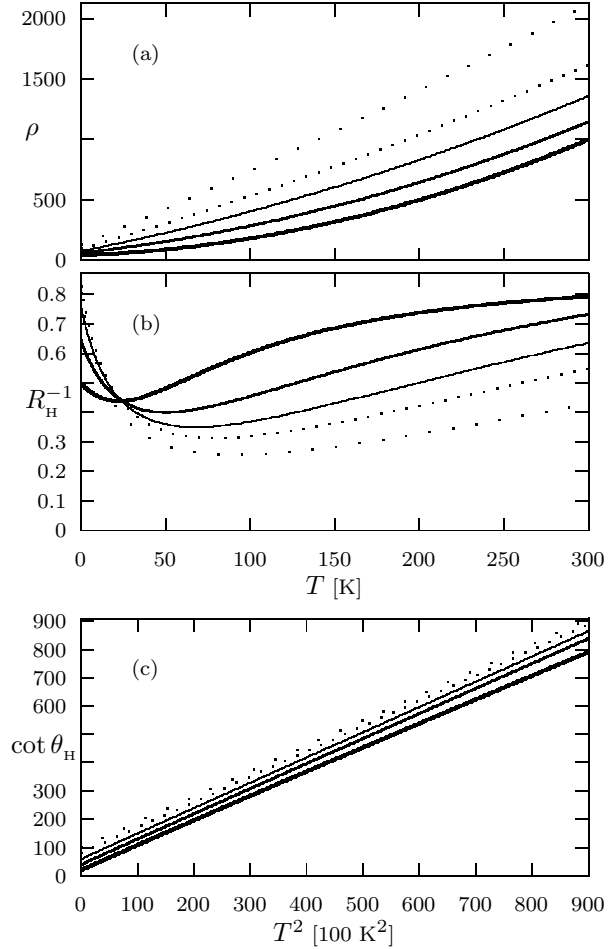


\setlength{\unitlength}{0.240900pt}
\ifx\plotpoint\undefined\newsavebox{\plotpoint}\fi
\sbox{\plotpoint}{\rule[-0.175pt]{0.350pt}{0.350pt}}%


\caption{The resistivity (a), inverse Hall constant (b), and
$\cot{\theta_{_{\rm H}}}$ (c), in arbitrary unit, for parameter values:
A=20,40,60,80,100; B=.01; C=.5,2,5,10,20; D=20,40,80,160,320;
N=1,1.3,1.8,2.5,3.4; Z=.01. The first value corresponds to the thickest
lines.} 
\label{F2}
\end{figure}

The transport quantities are then expressed as:
\begin{eqnarray}
\rho_x &\cong& {D+CT+A+BT^2 \over N}, \\ {1 \over \cot{\theta_{_{\rm
H}}}} &\cong& {Z \over D+CT} + {1 \over A+BT^2}. 
\end{eqnarray} 

\vskip -1.1truecm
\begin{figure}[t]
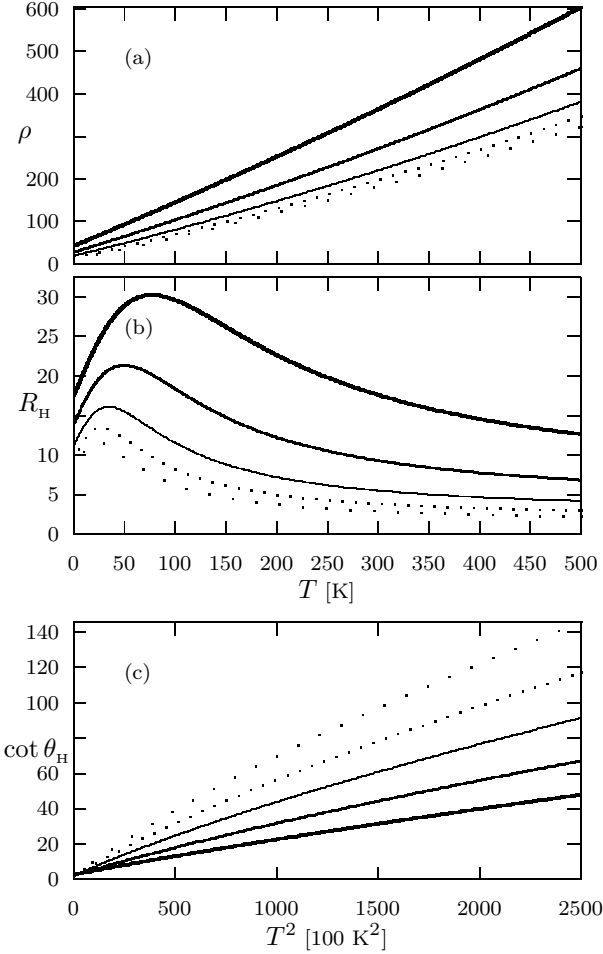


\setlength{\unitlength}{0.240900pt}
\ifx\plotpoint\undefined\newsavebox{\plotpoint}\fi
\sbox{\plotpoint}{\rule[-0.175pt]{0.350pt}{0.350pt}}%


\caption{The resistivity (a), Hall constant (b), and $\cot{\theta_{_{\rm
H}}}$ (c), in arbitrary unit, for parameter values: A=3,2.4,2,1.7,1.5;
B=.00025,.0004,.0006,.0008,.001; C=1,1.1,1.3,1.6,2; D=40;
N=1,1.5,2.2,3,4; Z=3. The first value corresponds to the thickest
lines.} 
\label{F3}
\end{figure}


These expressions reproduce the systematic behavior of the transport
quantities in different cuprates, except for the effect of the pseudogap
in underdoped cuprates, ignored here. Results for sets of parameters
corresponding to data in YBa$_2$Cu$_{3-x}$Zn$_x$O$_7$ [12],
Tl$_2$Ba$_2$CuO$_{6+\delta}$ [13], and La$_{2-x}$Sr$_x$CuO$_4$ [14], are
presented in Figs. 1, 2, and 3, respectively. 

The idea that $\rho$ and $\cot{\theta_{_{\rm H}}}$ are determined by
different scattering rates has been first suggested by Anderson [15],
and in his analysis the $T^2$ term is due to spinons. However, it has
been observed in ac Hall effect results [16] that the energy scale
corresponding to this term is of about $120\;$K. This energy is in
agreement with the very low energies of the stripons of our analysis,
and not with spinon energies (which are of tenths of an eV). 

\bigskip
\par\noindent{\bf 6.3 Thermoelectric Power}
\medskip

When both an electric field and a temperature gradient are present, 
one can express ${\bf j}^q$ and ${\bf j}^p$ as:
\begin{eqnarray}
{\bf j}^q &=& {{\rm e} \over T} \underline{\bf L}^{q(11)} \bfcalE^q +
\underline{\bf L}^{q(12)} \bfnabla({1 \over T}), \\
{\bf j}^p &=& {{\rm e} \over T} \underline{\bf L}^{p(11)} \bfcalE^p +
\underline{\bf L}^{p(12)} \bfnabla({1 \over T}).
\end{eqnarray} 
The thermoelectric power (TEP) is given by 
\begin{equation}
\underline{\bf S}=\bigg[{{\bf E} \over \bfnabla T}\bigg]_{{\bf j} = 0}. 
\end{equation}
By Eq. (7), the condition ${\bf j}=0$ for the evaluation of  the TEP 
becomes ${\bf j}^q\cong{\bf j}^p\cong 0$. 

Thus, by introducing
\begin{eqnarray}
\underline{\bf S}^q &=& \bigg[{\bfcalE^q \over \bfnabla T}\bigg]_{{\bf
j}^q = 0} = - {1 \over {\rm e}T} {\underline{\bf L}^{q(12)} \over
\underline{\bf L}^{q(11)}}, \\ \underline{\bf S}^p &=& \bigg[{\bfcalE^p
\over \bfnabla T}\bigg]_{{\bf j}^p = 0} = - {1 \over {\rm e}T}
{\underline{\bf L}^{p(12)} \over  \underline{\bf L}^{p(11)}}, 
\end{eqnarray}
and using Eqs. (11), (23), (24), we get that the TEP can be expressed
as: 
\begin{equation}
\underline{\bf S}={N^q\underline{\bf S}^q + N^p\underline{\bf S}^p \over
N^q + N^p}. 
\end{equation}

The QE bandwidths are close to an eV. Thus one gets 
\begin{equation}
S^q\propto T,
\end{equation}
similarly to the TEP in normal metals. On the other hand the stripon
bandwidth is of order $0.02\;$eV. Thus one expects $S^p$ to saturate at
$T \sim 200\;$K to the narrow-band result [17]: 
\begin{equation}
S^p={k_{_{\rm B}} \over {\rm e}} \ln{\bigg[{1-n^p \over n^p}\bigg]}, 
\end{equation}
where $n^p$ is the fractional occupation of the stripon band. 

This result for the TEP is consistent with the typical behavior observed
in the cuprates. Such a behavior has been parametrized as [18]: 
\begin{equation}
S=AT+{BT^{\alpha} \over (T+\Theta)^{\alpha}}. 
\end{equation} 
It was found [17,19] that $S^p=0$ (namely the stripon band is half full)
for slightly overdoped cuprates. 

The effect of the doping of a cuprate is [7] both to change the density
of the charged stripes within a CuO$_2$ plane, and to change the density
of carriers (stripons) within a charged stripe. It is the second type of
doping effect that changes $n^p$. overdoping is often limited because a
large density carriers in the charged stripes results in an increase of
Coulomb repulsion energy. 

\bigskip
\par\noindent{\bf 7. MECHANISM FOR HIGH-\pmb{$T_c$}\/}
\medskip

The coupling Hamiltonian ${\cal H}^{\prime}$ (3) provides a mechanism
for high-$T_c$. The pairing mechanism involves transitions between pair
states of QE's and stripons through the exchange of svivons. Such a
mechanism has similarities to the interband pair transition mechanism
proposed by Kondo [20]. 

The symmetry of the superconducting gap is strongly affected by the
symmetry of the QE-stripon coupling through ${\cal H}^{\prime}$, and is
thus related to the symmetry of the normal-state pseudogap (also 
determined by QE-stripon coupling). Similarity between the ${\bf 
k}$-space symmetries of these gaps has been observed [11].

A condition for superconductivity within the present approach is 
that the narrow stripon band maintains coherence between different 
stripe segments. Thus an upper limit for $T_c$ is determined by the
temperature where such coherence sets in.

When the stripon band is almost empty (or almost full), it can be
treated in the parabolic approximation, and it is characterized by the
distance ${\cal E}_{_{\rm F}}$ of the Fermi level from the bottom (top)
of the band at $T=0$. Using a two-dimensional approximation one can
express ${\cal E}_{_{\rm F}}$ as: 
\begin{equation}
{\cal E}_{_{\rm F}} = 2\pi\hbar^2 {n^* \over m^*}, 
\end{equation}
where $m^*$ in the stripons effective mass and $n^*$ is their density 
per unit area of a CuO$_2$ plane (note that the stripons are spinless).

Stripon coherence is energetically favorable at temperatures where there 
is a clear cut between occupied and unoccupied stripon band states,
which is of order of ${\cal E}_{_{\rm F}}/k_{_{\rm B}}$. And this
determines an upper limit for $T_c$: 
\begin{equation}
k_{_{\rm B}} T_c \lta {\cal E}_{_{\rm F}}.
\end{equation}

This result agrees with the ``Uemura plots'' [21] if the $n^*/m^*$ ratio
for stripons is approximately proportional to that for the supercurrent
carriers, appearing in the expression for the London penetration depth
(the supercurrent carriers are hybridized QE and stripon pairs). The
``boomerang-type'' behavior of the Uemura plots in overdoped cuprates
[22] is understood as a transition between a band-top and a band-bottom
behavior. 

\bigskip
\par\noindent{\bf 8. SUMMARY}
\medskip

The electronic structure of the high-$T_c$ cuprates has been studied on
the basis of both large-$U$ and small-$U$ orbitals. A striped structure
and three types of carriers are obtained; polaron-like stripons carrying
charge, QE's carrying charge and spin, and svivons carrying spin and
lattice distortion. 

Anomalous normal-state properties of the cuprates are understood, and
the systematic behavior of the resistivity, Hall constant, and
thermoelectric power is explained. The high-$T_c$ mechanism is based on
transitions between pair states of stripons and QE's through the
exchange of svivons. 




\bigskip
{\bf REFERENCES \hfill } \\

\vskip -0.15truecm
\baselineskip 10pt
\footnotesize
\noindent
1. \ Papers in {\it J.~Supercond.} {\bf 10}, \#4 (1997), and in this
issue. \\
2. \ O.~K.~Andersen {\it et al.}, {\it J.~Phys.~Chem.~Solids} {\bf 56}, 
1573 \\ \mbox{\ } \ \ \ (1995). \\
3. \ J.~Ashkenazi, {\it J.~Supercond.} {\bf 7}, 719 (1994); {\it ibid}
{\bf 8}, 559 \\ \mbox{\ } \ \ \ (1995); {\it ibid} {\bf 10}, 379 (1997).
\\ 
4. \ S.~E.~Barnes, {\it Adv.~Phys.} {\bf 30}, 801 (1980). \\
5. \ V.~J.~Emery, and S.~A.~Kivelson, {\it Physica C} {\bf 209}, 597 
\\ \mbox{\ } \ \ \ (1993). \\
6. \ A.~Bianconi {\it et al.}, {\it Phys.~Rev.~B} {\bf 54}, 12018, (1996); 
\\ \mbox{\ } \ \ \ {\it Phys.~Rev.~Lett.} {\bf 76} 3412 (1996). \\ 
7. \ J.~M.~Tranquada {\it et al.}, {\it Phys.~Rev.~B} {\bf 54}, 7489, (1996);
\\ \mbox{\ } \ \ \ {\it Phys.~Rev.~Lett.} {\bf 78}, 338 (1997). \\ 
8. \ C.~M.~Varma {\it et al.}, {\it Phys.~Rev.~Lett.} {\bf 63}, 1996
(1989). \\ 
9. \ D.~B.~Tanner, and T.~Timusk, {\it Physical Properties \\ \mbox{\ }
\ \ \ of High Temperature Superconductors III}, edited by \\ \mbox{\ } \
\ \ D.~M.~Ginsberg (World Scientific, 1992), p. 363. \\ 
10. M.~I.~Salkola, {\it et al.}, {\it Phys.~Rev.~Lett.} {\bf 77}, 155
(1996). \\ 
11. D.~S.~Marshall, {\it et al.}, {\it Phys.~Rev.~Lett.} {\bf 76}, 4841
(1996). \\ 
12. T.~R.~Chien, {\it et al.}, {\it Phys.~Rev.~Lett.} {\bf 67}, 2088
(1991). \\ 
13. Y.~Kubo and T.~Manako, {\it Physica C} {\bf 197}, 378 (1992). \\
14. H.~Takagi, {\it et al.}, {\it Phys.~Rev.~Lett.} {\bf 69}, 2975
(1992); \\ \mbox{\ } \ \ \ H.~Y.~Hwang, {\it et al.}, {\it ibid.} {\bf
72}, 2636 (1994). \\ 
15. P.~W.~Anderson, {\it Phys.~Rev.~Lett.} {\bf 67}, 2092 (1991). \\ 
16. H.~D.~Drew, S.~Wu, and H.-T.~S. Linh, preprint. \\
17. B.~Fisher, {\it et al.}, {\it J. Supercond.} {\bf 1}, 53 (1988);
J. Genossar, \\ \mbox{\ } \ \ \ {\it et al.}, {\it Physica C} {\bf
157}, 320 (1989). \\ 
18. S. Tanaka, {\it et al.}, {\it J.~Phys.~Soc.~Japan} {\bf 61}, 1271 
(1992). \\
19. K.~Matsuura, {\it et al.}, {\it Phys.~Rev.~B} {\bf 46}, 11923
(1992); \\ \mbox{\ } \ \ \ \ S.~D.~Obertelli, {\it et al.}, {\it ibid.},
p. 14928; C.~K.~Subramaniam, \\ \mbox{\ } \ \ \ {\it et al.}, {\it
Physica C} {\bf 203}, 298 (1992). \\ 
20. J.~Kondo, {\it Prog.~Theor.~Phys.} {\bf 29}, 1 (1963). \\ 
21. Y.~J.~Uemura, {\it et al.}, {\it Phys.~Rev.~Lett.} {\bf 62}, 2317
(1989). \\
22. Ch. Niedermayer, {\it et al.}, {\it Phys.~Rev.~Lett.}~{\bf 71}, 1764
(1993). \\
 
\end{document}